\documentclass[journal]{IEEEtran}
\usepackage[utf8]{inputenc}
\usepackage{url}
\usepackage[bottom]{footmisc}
\usepackage{hyperref}

\usepackage{xcolor}
\usepackage{graphicx}
\usepackage{amsmath}

\ifCLASSINFOpdf
\else
\fi
\usepackage{stackrel}
\begin{document}
%
\title{Desmoking laparoscopy surgery images using an image-to-image translation guided by an embedded dark channel}

\author{
Sebastián Salazar-Colores,
Hugo Alberto-Moreno,
César Javier Ortiz-Echeverri,
and~Gerardo Flores
\thanks{Sebastián Salazar-Colores,
Hugo Alberto-Moreno and
Gerardo Flores are with Optical Research Center  (CIO),
Lomas del Bosque 115, Lomas del Campestre, 37150 León, Gto (https://www.cio.mx).}

\thanks{César Javier Ortiz-Echeverri is with Universidad Autónoma de Querétaro,
Facultad de informática, Av. de las Ciencias S/N, 76230 Juriquilla, Qro. (https://www.uaq.mx).}

}

\maketitle

\begin{abstract}
In laparoscopic surgery, the visibility in the image can be severely degraded by the smoke caused by the $CO_2$ injection, and dissection tools, thus reducing the visibility of organs and tissues. This lack of visibility increases the surgery time and even the probability of mistakes conducted by the surgeon, then producing negative consequences on the patient's health. In this paper, a novel computational approach to remove the smoke effects is introduced. The proposed method is based on an image-to-image conditional generative adversarial network in which a dark channel is used as an embedded guide mask. Obtained experimental results are evaluated and compared quantitatively with other desmoking and dehazing state-of-art methods using the metrics of the Peak Signal-to-Noise Ratio (PSNR) and Structural Similarity (SSIM) index. Based on these metrics, it is found that the proposed method has improved performance compared to the state-of-the-art. Moreover, the processing time required by our method is 92 frames per second, and thus, it can be applied in a real-time medical system trough an embedded device.
\end{abstract}


%
\IEEEpeerreviewmaketitle

\section{Introduction}
\label{sec:introduction}
Laparoscopic surgery involves the insertion of a camera through small incisions in the patient's body, where an inert gas $CO_2$ is injected in order to expand the abdomen to accommodate the surgical instruments \cite{sauerland2010}. The use of $CO_2$ gas and the dissection of tissues during surgery decrease the organs’ visibility, thus reducing medical and robotic support systems' performance \cite{gu2015virtual,hahn2017removal}. On the other hand, clinical studies have shown that digital smoke removal reduces surgery operative time and surgeons anxiety during surgical operation \cite{gu2015virtual, Lawrentschuk}. With that in mind, recently researchers have been conducted works to reduce, or even avoid the smoke effects. Commonly, the smoke removal process is performed with medical instruments \cite{goodson1988laser}. However, such methods are costly and impractical, and for those reasons recently image processing approaches and deep learning techniques have raised as a solution to such a problem
\cite{kotwal2016joint,baid2017joint,luo2017vision,wang2018smoke, shin2019radiance, gu2015virtual,zhu2015fast,Tchaka2017,wang2018variational,bolkar2018deep,Sidorov2019,wang2019multiscale,Vishal2019,chen2018unsupervised}.
\begin{figure}[h]
    \centering
    \includegraphics[width=0.45\textwidth]{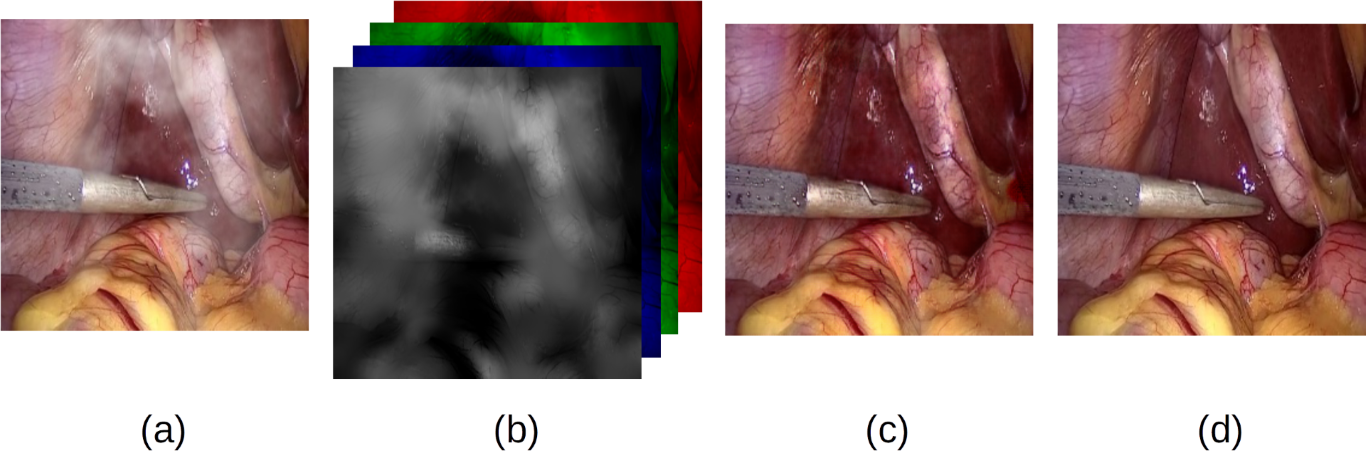}
    \caption{The proposed approach for solving the digital smoke removal in laparoscopic surgery: (a) Input image with simulated smoke; (b) input image with its corresponding embedded dark channel; (c) output proposed model; and (d) ground-truth.}
    \label{fig:01}
\end{figure}

The literature that addresses the solution of the digital smoke removal problem can be divided into three main approaches: a) traditional image processing techniques \cite{kotwal2016joint,baid2017joint,luo2017vision,wang2018smoke, shin2019radiance}; b) physics-model-based methods, especially the atmospheric scattering model and the dark channel prior (DCP) \cite{gu2015virtual,zhu2015fast,Tchaka2017,wang2018variational}; these methods are inspired by the excellent results achieved in dehazing tasks \cite{xu2015review}; and c) where deep learning methods based on convolutional neural networks and generative networks are implemented \cite{bolkar2018deep,chen2018unsupervised,Sidorov2019,wang2019multiscale,Vishal2019,vishal2020unsupervised}. The first two approaches have been widely studied in the literature, while the third approach is barely being explored in the last two years. The contribution of this work lies in the latter approach, where a novel generative desmoking method is introduced. In particular, our method is based on a conditional generative adversarial network (cGAN) in which a guide mask based on dark channel is embedded. In Fig. \ref{fig:01} an example of the proposed approach in the present work is shown.

\subsection{Related work}
The problem of digital smoke removal seeing from the perspective of classical image processing techniques has been investigated in the following works. In \cite{kotwal2016joint} and \cite{baid2017joint}, joint image desmoking and denoising of laparoscopic images is formulated as a Bayesian inference problem. In \cite{luo2017vision}, the method first recovers the visibility and enhances the contrast of hazy images, then luminance of enhanced images are fused in the gradient domain by solving the Poisson equation in the frequency domain. In \cite{wang2018smoke}, a physical model of gas scattering and Lagrangian methods were used to remove unwanted smoke components. In \cite{shin2019radiance}, the authors propose an algorithm with the aim to remove the dehazing artifacts by optimizing the transmission map; the method is based on the estimated reflectance map and the structure-guided transmission map $\ell _0$.

The physics-model-based approach includes the methods based on the atmospheric scattering model, and dark channel prior (DCP) to estimate the transmission map and atmospheric light. In this regard, the state-of-the-art is as follows. In \cite{zhu2015fast}, the authors present a color attenuation prior to haze removal. Their method is based on a linear model that learns the parameters by means of a supervised learning method. In \cite{Tchaka2017}, it was proposed an adapted dark channel prior method combined with histogram equalization to remove smoke artifacts, and thus recover the radiance image. The authors of \cite{wang2018variational} propose a method based on the assumption that the smoke veil has low contrast and low inter-channel differences; a cost function is defined based on this prior knowledge and is solved using an augmented Lagrangian method. In the technological aspect, the authors of \cite{gu2015virtual} have implemented a digital system based on the dark channel prior. 

Since the smoke patterns are complex and hard to categorize, the aforementioned couple of approaches generally produces inconsistent results, altering and saturating colors in no-smoke regions. To overcome this, artificial-intelligence-based solutions have shown superior results in dehazing-like problems, in which the haze effect must be eliminated in the image of interest. The state-of-the-art in this particular subject is as follows. In \cite{bolkar2018deep} a deep neural network is implemented in real-time to enhance the quality of surgery video frames trying to eliminate the smoke presented in the image. The authors of \cite {Sidorov2019} propose an unsupervised image-to-image translation with a generative adversarial network (GAN) architecture; they use a perceptual image quality metric for the loss function. In \cite{wang2019multiscale}, a convolutional neural network (CNN) takes the smoke image and its pyramidal decomposition as inputs, and then, it directly outputs a smoke-free image without relying on any physical model, neither estimation of intermediate parameters. In \cite{Vishal2019}, it is implemented an end-to-end network called Cycle-Desmoke. This network uses a generator architecture and two loss functions: a guided-unsharp upsample loss function, and an adversarial and cycle-consistency loss function.

In \cite{chen2018unsupervised}, it is implemented an unsupervised framework for learning smoke removal that uses a fully convolutional encoder-decoder network to generate the same size de-smoked image. In \cite{vishal2020unsupervised}, an unsupervised deep learning relied on GAN converts laparoscopic images from smoke domain to smoke-free domain. The network comprises a generator architecture endowed with an encoder-decoder structure composed of multi-scale feature extraction at each encoder block. Then, it obtains a robust deep representation map with the aim to reduce the image's smoke component.


\subsection{Contribution}
Inspired by deep learning approaches where embedded image masks provide additional information to the CNN \cite{ren2019mask, yildirim2018disentangling, gu2019mask}, and as dark channel intensity is a good indicator of smoke level presence \cite{xu2015review}, we propose a combination of a conditional generative adversarial network and the embedded dark channel mask. The image mask gives concentration smoke data to the conditional generative adversarial network to identify the restoration level needed in each image region. Obtained results in synthetic images show that the proposed approach is able to remove local smoke and recover realistic tissue colors without affect non-smoke areas. According to the peak signal-to-noise ratio (PSNR) and structural similarity (SSIM) index, our method outperforms seven state-of-the-art methods, including the image-to-image conditional generative model with no embedded mask (Pix2Pix). Obtained results can be reproduced with the source code and trained
models available at our GitHub site presented in Section \ref{sec:res}.

The present work is structured as follows: In Section \ref{sec:meth} it is presented the key idea around our proposed method, including the foundations that support the method. Section \ref{sec:res} explains the setup used in the design of the experiment and the obtained results, just as the comparative Tables and Figures with respect to six state-of-the-art methods. Finally, conclusions and a perspective of future research are presented in Section \ref{sec:conc}.

\section{Proposed method}\label{sec:meth}
In this Section the motivation, architecture, loss function, and theoretical background of the proposed approach are presented.
\subsection{Overall procedure of the proposed method}
The proposed method is inspired by the following two approaches:
\begin{itemize}
    \item A convolutional neural network oriented to classify tasks that receive as inputs an RGB image and its corresponding salience map (S). A salience map S is an image that tries to predict human fixations. The obtained results of this method show that performance improves when meaningful data are incorporated \cite{Murabito2018}.
    \item We use the fact that object features such as shape, color, or even pose, can be embedded from input masks in generative adversarial networks. This fact has been demonstrated in \cite{ren2019mask, yildirim2018disentangling, gu2019mask}.
 \end{itemize}
 
Since in an image there exists a relation between the dark channel and the presence of smoke, the key idea of the present work is that by embedding a dark channel mask into the input of a cGAN, it is possible to obtain significant information that can improve the desmoking performance in laparoscopic images. This is explained next.

\subsection{The relation between the dark channel and laparoscopic images}
\label{2b}
The dark channel ($I^\textrm{dark}$) for each pixel $(x,y)$ is defined as
\begin{equation}\label{eq1}
    I^\textrm{dark}(x,y) = \min_{c\in \{R,G,B\}}{\left({\min_{z\in \Omega(x,y)}} I^c(z)\right)} ,
\end{equation}
where $\Omega(x,y)$ is a kernel, usually squared, centered in the $(x,y)$ position; $I^c(z)$ are the elements of the smoked laparoscopic image $I$ in the positions $z\in\Omega(x,y)$; and $c$ represent each channel component of the $RGB$. In \cite{he2010single}, it was observed that in the vast majority of regions without haze or smoke the dark channel tends to have low values, this means that $I^\textrm{dark}\to 0$ holds. This statistical fact was named as \textit{dark channel prior}.

\begin{figure}[h]
    \centering
    \includegraphics[width=0.48\textwidth]{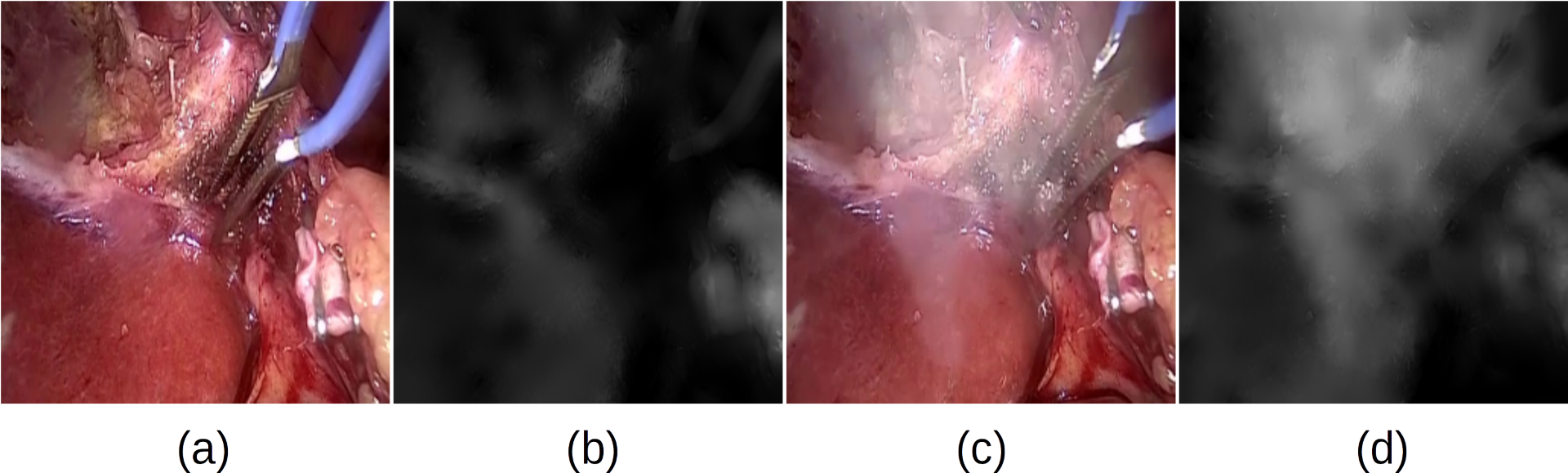}
    \caption{Dark channel in clear and smoked images: (a) clear laparoscopic image; (b) dark channel of (a); (c) smoked laparoscopic image; and (d) dark channel of (c).}
    \label{dcd}
\end{figure}

Since in laparoscopic images, the tissues and organs are commonly bright and colorful, similar to outdoor natural images, the dark channel is still valid to discriminate between smoke and clear images. In Fig. \ref{dcd} it is shown the dark channel map of laparoscopic images in two conditions: when it is free of smoke, and when it is smoked. It is visible that smoke and clear regions tend to have high and low-intensity values, respectively.

\subsection{Obtainment of refined dark channel}
\begin{figure}[h]
    \centering
    \includegraphics[width=0.48\textwidth]{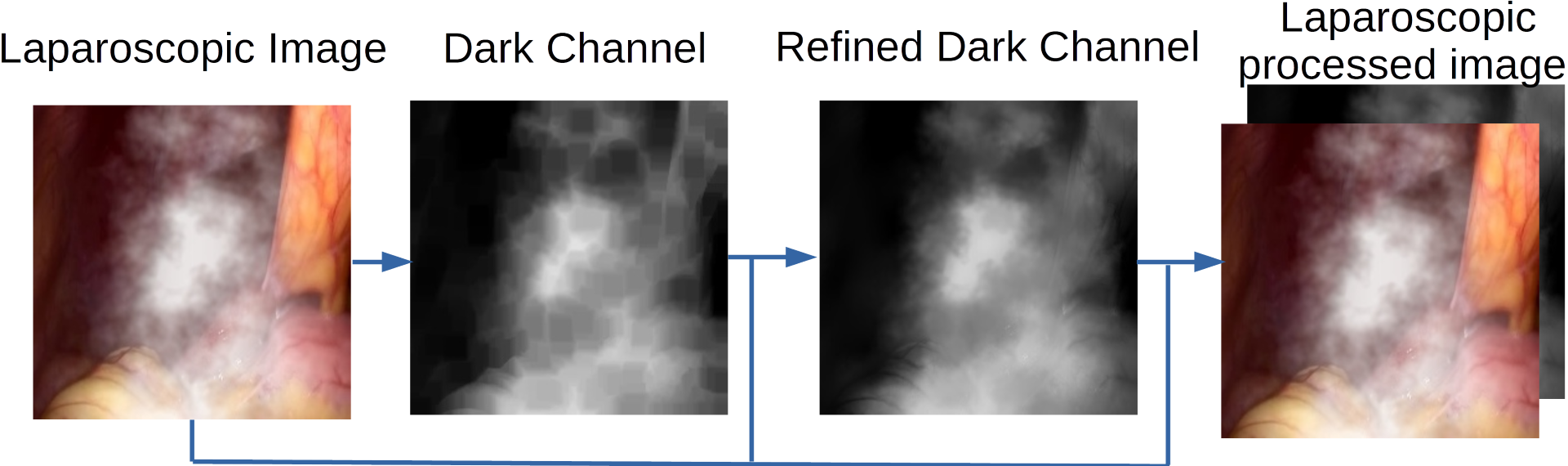}
    \caption{Process of computing, refinement, and embedding of the dark channel into the smoked laparoscopic image.}
    \label{add-dc}
\end{figure}

Since in dark channel computation the implemented kernels $\Omega(x,y)$ are squared patches, the results of the dark channel are not exactly aligned to the corresponding image $I$, and hence is not spatial accurate. For solving that, it is mandatory to integrate a \textit{refinement stage} for the dark channel maps. For that, several methods have been proposed such as \cite{pang2011improved} and \cite{Lee2016}. In order to obtain an accurate refined dark channel map showed in \eqref{eq1}, in this paper we apply a \textit{guided filter} to $I^\textrm{dark}$, which is an edge-preserving smoothing filter based on a model linearized in a region\cite{10.1007/978-3-642-15549-9_1}. Such a filter is given as follows
\begin{equation}\label{eq:filter}
    I^\textrm{dark}_\textrm{ref}(z) = a{(x,y)} I(z) + b{(x,y)}, \forall z \in \Omega(x,y),
\end{equation}
where $I^\textrm{dark}_\textrm{ref}$ is the filtering output dark channel; $I$ is the guidance image; and $z$ is the position of a pixel in the local squared window $\Omega$ of size $s\times s$ and centered in $(x,y)$. The $a(x,y)$ and $b(x,y)$ parameters from linear model \eqref{eq:filter} are defined as
\begin{subequations} \label{eq:min}
     \begin{align}
	 a(x,y) &= \frac{\frac{1}{|\Omega|}\sum_{(z)\in\Omega(x,y)}I{(z)}I^\textrm{dark}{(z)}-\mu{(x,y)}\overline{{I}^\textrm{dark}{(x,y)}}}{\sigma{(x,y)}^2+\epsilon}, \label{eq:min_a} \\
	b(x,y) &= \overline{{I}^\textrm{dark}{(x,y)}}-a{(x,y)}\mu{(x,y)}, \label{eq:min_b}
	  \end{align}
\end{subequations}
where $\mu{(x,y)}$ and $\sigma{(x,y)}$ are the mean and variance of $I$ in $\Omega{(x,y)}$; $\overline{{I}^\textrm{dark}{(x,y)}}$ is the mean
of ${I}^\textrm{dark}{(x,y)}$ in $\Omega_{k}$; and $\epsilon$ is parameter that regulates the smoothness degree. The computed and refined dark channel $I^\textrm{dark}_\textrm{ref}(z)$ is stacked into the smoked laparoscopic image $I$, as it is shown in Fig. \ref{add-dc}.

Once the refined dark channel is embedded, the processed images are used as input of the proposed cGAN. This is explained next.

\subsection{Conditional Generative Adversarial Networks}
Generative adversarial networks (GANs) \cite{goodfellow2014generative} are unsupervised generative models that learn a mapping $\mathbf{G}:r \rightarrow o$, where $r$ is a random noise vector, and $o$ is an output image. Such a mapping is implemented by a system of two competitor neural networks: a \textit{generator} (G) and a \textit{discriminator} (D). The generator is commonly formed by convolutional and deconvolutional layers, which are designed to produce outputs that cannot be distinguished from real images. The discriminator is given by a CNN trained to perform its best in detecting the counterfeits made by \emph{G}. 

\begin{figure}[h]
    \centering
    \includegraphics[width=0.35\textwidth]{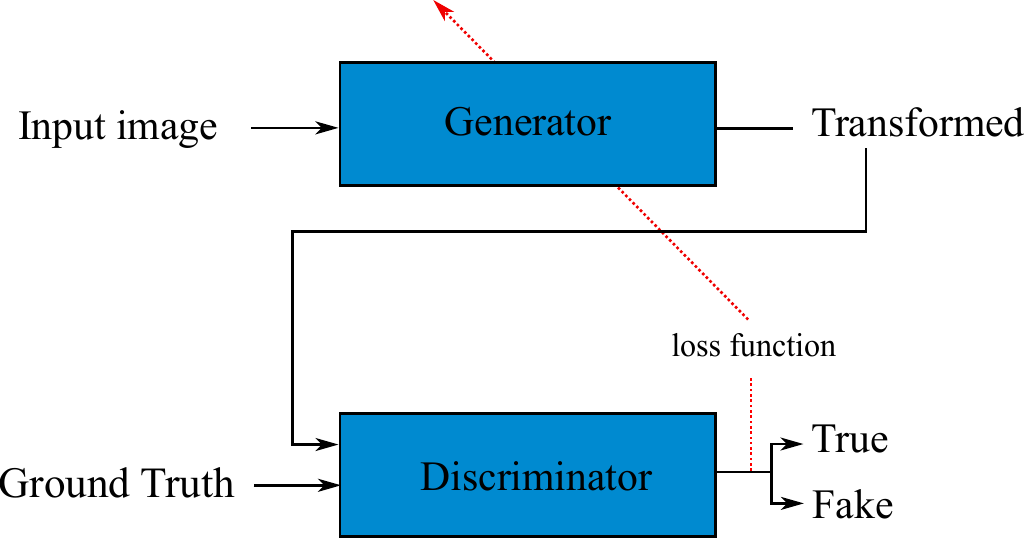}
    \caption{General scheme of cGAN architecture.}
    \label{GAN1}
\end{figure}

The conditional generative adversarial networks (cGAN's), unlike the GAN, additionally uses a conditional input vector $i$ in its mapping, i.e. $\mathbf{G}:i,r \rightarrow o$ \cite{isola2017image}. It is highlighted that in the majority of the tasks that can be reduced to an image-to-image translation problem, the cGAN approach is currently the state-of-the-art \cite{karras2019style, zhang2019image, wang2018video, zhang2017adversarial}. In Fig. \ref{GAN1} it is depicted a general scheme of the used network architecture.
%
%
\subsection{Architecture}
\begin{figure}[!ht]
    \centering
    \includegraphics[width=0.47\textwidth]{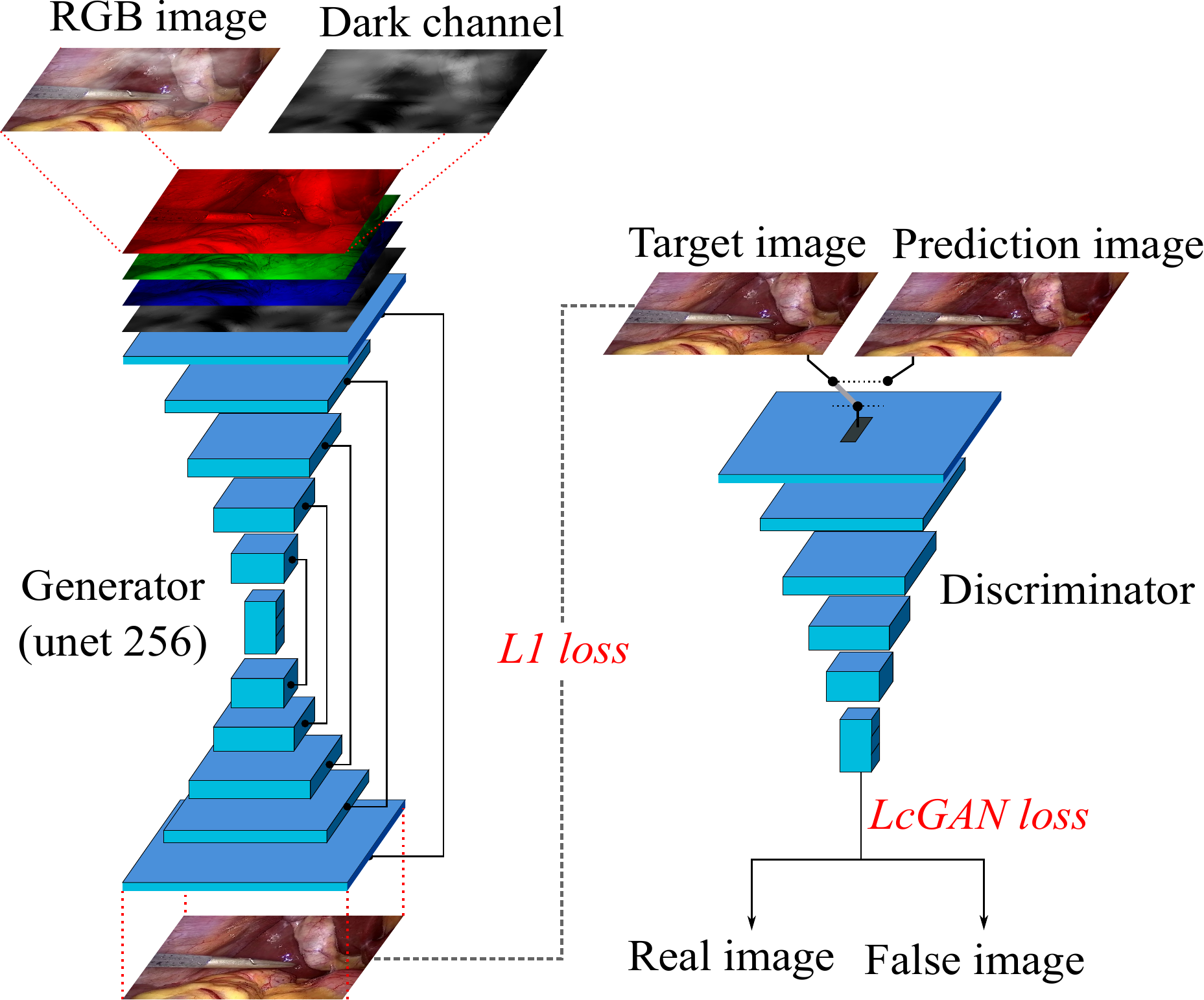}
    \caption{A scheme of the proposed approach. The input is the refined dark channel stacked into a smoked RGB image. Such RGB image is then processed by the generator to obtain the de-smoked laparoscopic images.}
    \label{GAN2}
\end{figure}

The proposed method is based on the pix2pix architecture presented in \cite{isola2017image}. The generator is based on a U-Net architecture with images of input of $256^2$ pixels. U-Net’s architecture is similar to an Auto-Encoder. Both U-Net and Auto-Encoder have two networks: encoder and decoder. The difference between \textit{U-Net} and \textit{Auto-Encoder} network is that \textit{U-Net} skips connections between encoder layers and decoder layers, while \textit{Auto-Encoder} does not. Both architectures have a stage where the features of the inputs are described, the latent space and bottleneck layer, respectively. In Fig. \ref{GAN2} it is depicted the proposed cGAN architecture, while in Table \ref{tabla2} and Table \ref{tabla1} the hyper-parameters of \emph{D} and \emph{G} are detailed.

\begin{center}
\begin{table}[!ht]
\renewcommand{\arraystretch}{1.8}
\caption{Architecture of employed Discriminator \emph{D}.}
\normalsize
\center
\resizebox{0.48\textwidth}{!}{
\begin{tabular}{cccclc}
\hline
\textbf{Layer} & \textbf{Conv} & \textbf{kernel} & \textbf{Strides} & \textbf{Definition}                                              & \textbf{Size}         \\ \hline\hline
1             & 64          & 4      & 2      & (Conv -\textgreater BatchNorm -\textgreater Leaky ReLU) & (128, 128, 64) \\ \hline
2             & 128         & 4      & 2      & (Conv -\textgreater BatchNorm -\textgreater Leaky ReLU) & (64, 64, 128)  \\ \hline
3             & 256         & 4      & 2      & (Conv -\textgreater BatchNorm -\textgreater Leaky ReLU) & (32, 32, 256)  \\ \hline
4             & 0           & 0      & 0      & (ZeroPadding2D)                                         & (34, 34, 256)  \\ \hline
5             & 512         & 4      & 1      & (Conv )                                                 & (31, 31, 512)  \\ \hline
6             & 0           & 0      & 0      & (BatchNorm -\textgreater Leaky ReLU-\textgreater ZeroPadding)        & (33, 33, 512)  \\ \hline
7             & 1           & 4      & 1      & (Conv)                                                  & (30, 30, 1)    \\ \hline
\end{tabular}}
\label{tabla2}
\end{table}
\end{center}

\begin{center}
\begin{table}[h]
\renewcommand{\arraystretch}{1.4}
\caption{Architecture of the generator \emph{G}. Capital letter $C$ denotes (Convolution -\textgreater BatchNorm -\textgreater Leaky ReLU); $CTD$ denotes (Deconvolution -\textgreater BatchNorm-\textgreater{}ReLU-\textgreater{}Dropout rate $50\%$); and $CT$ denotes (Deconvolution -\textgreater BatchNorm -\textgreater ReLU).}
\centering
\normalsize
\resizebox{0.48\textwidth}{!}{
\begin{tabular}{ccccccc}
\hline
\textbf{Layer} & \textbf{Conv} & \textbf{Kernel} & \textbf{Strides} & \textbf{Definition}& \textbf{Size}& \textbf{Skip connection} \\ \hline\hline
1  & 64 & 4 & 2 & C             & (128, 128, 64)     & $\rightarrow$ 16 \\ \hline
2  & 128& 4 & 2 & C             & (64, 64, 128)      & $\rightarrow$ 15 \\ \hline
3  & 256& 4 & 2 & C             & (32, 32, 256)      & $\rightarrow$ 14 \\ \hline
4  & 512& 4 & 2 & C             & (16, 16, 512)      & $\rightarrow$ 13 \\ \hline
5  & 512& 4 & 2 & C             & (8, 8, 512)        & $\rightarrow$ 12 \\ \hline
6  & 512& 4 & 2 & C             & (4, 4, 512)        & $\rightarrow$ 11 \\ \hline
7  & 512& 4 & 2 & C             & (2, 2, 512)        & $\rightarrow$ 10 \\ \hline
8  & 512& 4 & 2 & C             & (1, 1, 512)        &    n/a              \\ \hline
9  & 512& 4 & 2 & CTD           & (1, 1, 1024)       &    n/a                \\ \hline
10 & 1024& 4 & 2& CTD           & (2,2, 1024)        & 7 $\rightarrow$  \\ \hline
11 & 1024& 4 & 2& CTD           & (4,4, 1024)        & 6 $\rightarrow$  \\ \hline
12 & 1024& 4 & 2& CT            & (8, 8, 1024)       & 5 $\rightarrow$  \\ \hline
13 & 1024& 4 & 2& CT            & (16, 16, 1024)     & 4 $\rightarrow$  \\ \hline
14 & 512& 4  & 2& CT            & (32, 32, 512)      & 3 $\rightarrow$  \\ \hline
15 & 256& 4  & 2& CT            & (64, 64, 256)      & 2 $\rightarrow$  \\ \hline
16 & 128& 4 & 2 & CT            & (128, 128, 128)    & 1 $\rightarrow$  \\ \hline
17 & n/a     & 4 & 2 & tanh          & (256, 256, 3)      &   n/a   \\ \hline
\end{tabular}}
\label{tabla1}
\end{table}
\end{center}
%
%
\subsection{Objective function}
The loss function used in this work is based on a combination of a cGAN's objective function and a $L1$ loss function. This is described next.
\subsubsection{cGAN's objective function}
As in \cite{isola2017image}, the objective function of the cGAN is given as
\begin{equation}
    \mathcal{L}_{cGAN}(G,D)=\log D(I,J)+\log(1-D(I,G(I,r))),
\end{equation}
where $G$ is the generator; $D$ is the discriminator; $I$ is the smoked image; $J$ is the target image; $r$ is a random noise vector; and $G(I,r)$ are the restored image.
\subsubsection{Modification of the L1 loss function}
Since the laparoscopic images are only defined in the RGB channels, the output value of the fourth channel is only used as a guide, and it is not being relevant for the network performance. Then, we propose that the $L1$ loss function be focused only on the RGB channels, i.e.
\begin{equation}
     \mathcal{L} _{L1}(G)=\sum_{c\in R,G,B} \sum_{x,y}||J^c-[G(I,z)]^c||,
\end{equation}
where $(x, y)$ represent the pixel positions. Remember that $c$ represent each channel component of the $RGB$.

The final objective function used in this work is based on the previous loss functions and it is given by
\begin{equation}
    G*=\arg \min_{G}\max_{D}\mathcal{L}_{cGAN}(G,D)+\lambda \mathcal{L}_{L1}(G),
\end{equation}
where $\lambda>0$ is the weight of $L1$ loss function.
%
\section{Experimental results}\label{sec:res}
In this Section information related to the setup of the datasets, the architecture, and the realized comparison, as well as the obtained quantitative and qualitative results are shown.

\subsection{Setup}
The experiments were implemented in a Ryzen Threadripper processor with 128 GB of RAM memory and a graphic card Nvidia RTX 2080 Ti. The used operating system is Linux Ubuntu 18.10 endowed with Python 3.7, OpenCV libraries, and Pytorch 1.4.0 framework. Additionally, for comparison purposes, we have used Matlab R2019a to run some state-of-the-art methods.

\subsection{Datasets}\label{database}
\begin{figure}[!ht]
 \centering
    \includegraphics[width=0.45\textwidth]{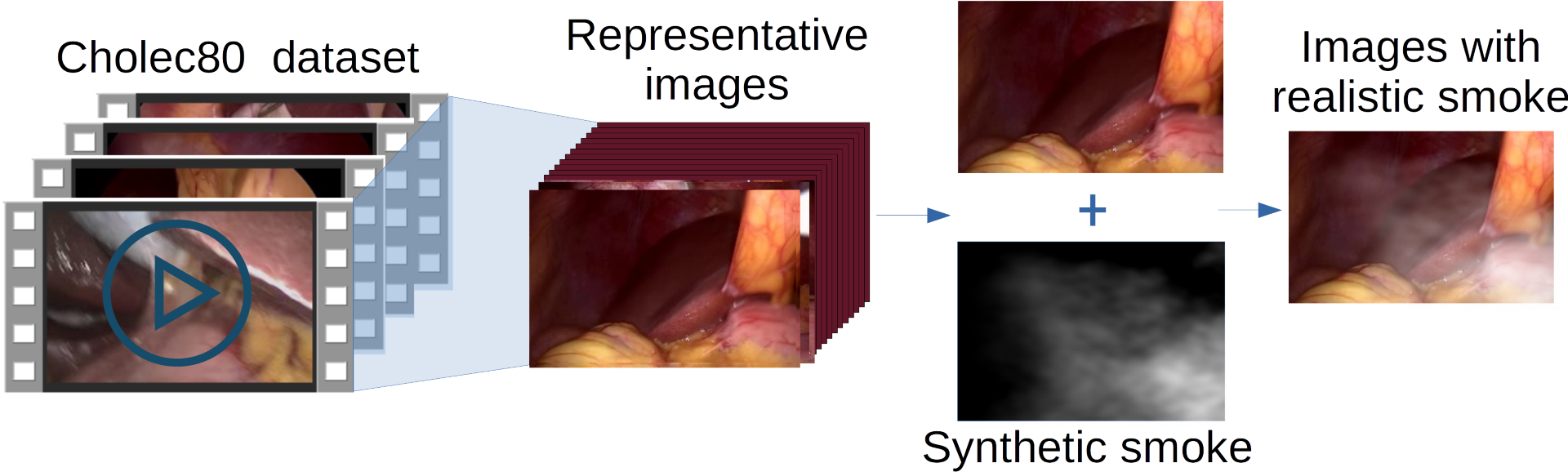}
    \caption{Process to generate the dataset.}
    \label{data}   

\end{figure}

As it is depicted in Fig. \ref{data}, the images used in this work were acquired from videos from Cholec80 dataset \cite{twinanda2016endonet} that contain $80$ videos of cholecystectomy surgeries performed by $13$ surgeons. From the available videos, two datasets were created: a train, and a test dataset. To generate the training dataset $20,000$ representative images without smoke were extracted from $50$ of the available videos. Synthetic random smoke was added as it is described in Section \ref{smoke_sim}. 

Similarly, to measure and compare the performance with the state-of-art methods, a test dataset was created with $2,398$ representative images extracted from $5$ of the aforementioned videos. It is highlighted that the videos used in the test dataset are different from those used in the training dataset. 

\subsection{Smoke simulation}
\label{smoke_sim}
A realistic simulation of heterogeneous smoke is important to train and test the developed model. In this work, smoke was simulated using Python-Clouds library\footnote{\href{https://github.com/SquidDev/Python-Clouds}{Link to Python-Clouds library}}, which uses Perlin noise as a base. Each simulation was saved in a normalized synthetic smoke mask $m(x,y)$. Smoke was added to the laparoscopic images using the scattering atmospheric model
\begin{equation}\label{addsmk}
 I^{c}\left ( x,y \right ) = J^{c}\left ( x,y \right )t(x,y)+(1-t(x,y))A^{c},
\end{equation}
the transmission map $t(x,y)$ is computed as the complement of $m(x,y)$ as follows
\begin{equation}\label{transmask}
t(x,y)= l(1-m(x,y)),
\end{equation}
where $c$ represents each of the $RGB$ channels; $l :(0\leq l\leq 1)$ is the intensity value; $A$ is a normalized $RGB$ atmospheric light color; and recall that $J(x,y)$ is a clear laparoscopic image, and $I(x,y)$ is the image with simulated smoke with dimensions of $256\times256$ pixels.
\subsection{Training}
The pix2pix architecture in which our method is based, was trained using a synthetic dataset. In order to compare the effectiveness of our mask embedding approach, the hyperparameters were selected the same in the original pix2pix and the proposed method. We used a batch size of $16$, a learning rate of $0.0002$, and a resolution of $256 \times 256$ pixels for the input and output images. Weights were initialized using a Gaussian distribution with zero mean and a standard deviation of $0.02$. Adaptable Momentum (ADAM) was used as an optimization function. Both networks were trained for $50$ epochs. In particular, epoch $35$ showed the best performance in our approach, while epoch $15$ showed the best for pix2pix. Pytorch framework was used in each model. The models were trained on two NVIDIA RTX 2080 Ti during one day. More details can be found in our source code \footnote{\href{https://github.com/ssalazarcolores/Desmoking-laparoscopy-surgery-images-using-an-image-to-image-translation-guided-by-an-embedded-Dark-}{Link to our code via github}}.

\subsection{Dark channel embedded mask's effect}
\begin{figure}[!ht]
    \centering
    \includegraphics[width=0.4\textwidth]{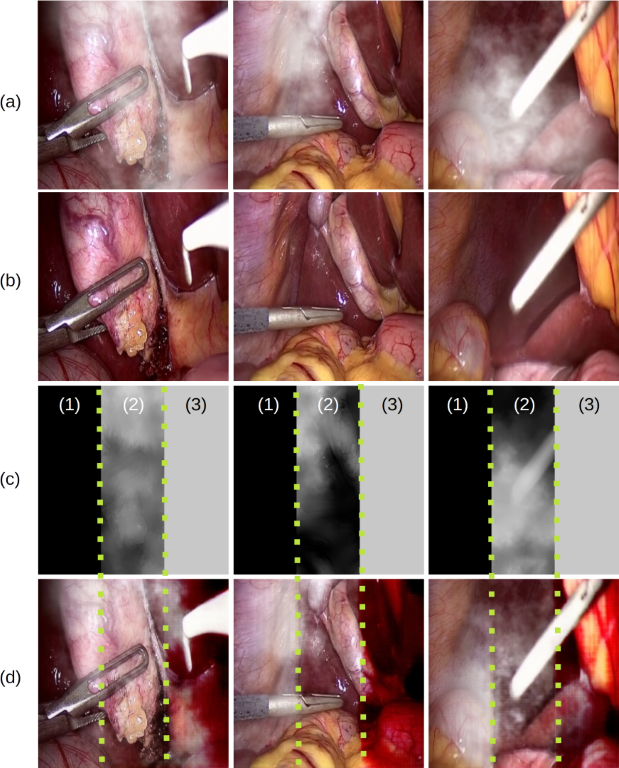}
    \caption{Effect of the dark channel-mask on image restoration. (a) Input image with synthetic smoke; (b) ground-truth; (c) dark channel-mask with manipulated values; and (d) output.}
    \label{fig8}   
\end{figure}

Fig. \ref{fig8} shows the decisive influence of the dark channel embedded mask in our trained model. In Fig. \ref{fig8}d the dark channel masks were manipulated and divided into three regions: low, original, and high dark channel values (from left to right in the figure). As it is expected, for the relation between the dark channel and laparoscopic images, when the dark channel has low values, the restoration tends to be imperceptible. In contrast, when dark channel values are high, the performed restoration saturates the pixels.

\subsection{Comparison}
In order to analyze the performance of the compared methods under different smoke intensities, three smoke levels were created: low, medium, and high. For that, it has been chosen three grades of intensity values, i.e., three sets in which the parameter $l$ in \eqref{transmask} can take values: $[0.40,  0.55]$, $[0.6, 0.75]$, and $[0.77. 0.92]$, corresponding to low, medium, and high smoke intensities, respectively. The particular value of $l$ is randomly chosen, provided that such a value belongs to any of the previously defined sets.

To demonstrate the capabilities of our method, we compare it against seven state-of-the-art methods. The seven compared methods can be divided into three groups: general-purpose dehazing methods, desmoking laparoscopic methods, and the pix2pix medhod. Next, we detail each of these methods.
\begin{itemize}
\item In the first group, it can be mentioned three single image-dehazing-general-purpose methods: the work of \cite{pang2011improved}, in which the authors combine dark channel prior and a guided image filtering; \cite{zhu2015fast} which uses a linear color attenuation prior based on the difference between brightness and saturation of pixels; and \cite{shin2019radiance}, where the authors presented optimization-based methods that combine radiance and reflectance components with additional refinement, all these using a guided-structure filter. 
\item The second group is three specialized desmoking laparoscopic methods: the work of \cite{bolkar2018deep}, where it is presented a convolutional architecture with multi-scale kernels; \cite{chen2018unsupervised} where the authors use an unsupervised learning approach based on U-Net structure; and \cite{wang2019multiscale} in which an approach based on CNN with a Laplacian image pyramid decomposition input strategy is used. 
\item The third group is given by the cGAN pix2pix approach (Isola et al.) \cite{isola2017image}. The comparison with such an approach permits to observe the impact of our proposed dark channel embedded mask.
\end{itemize}
\subsection{Qualitative comparison}
\begin{figure*}
    \centering
    \includegraphics[width=0.98\textwidth]{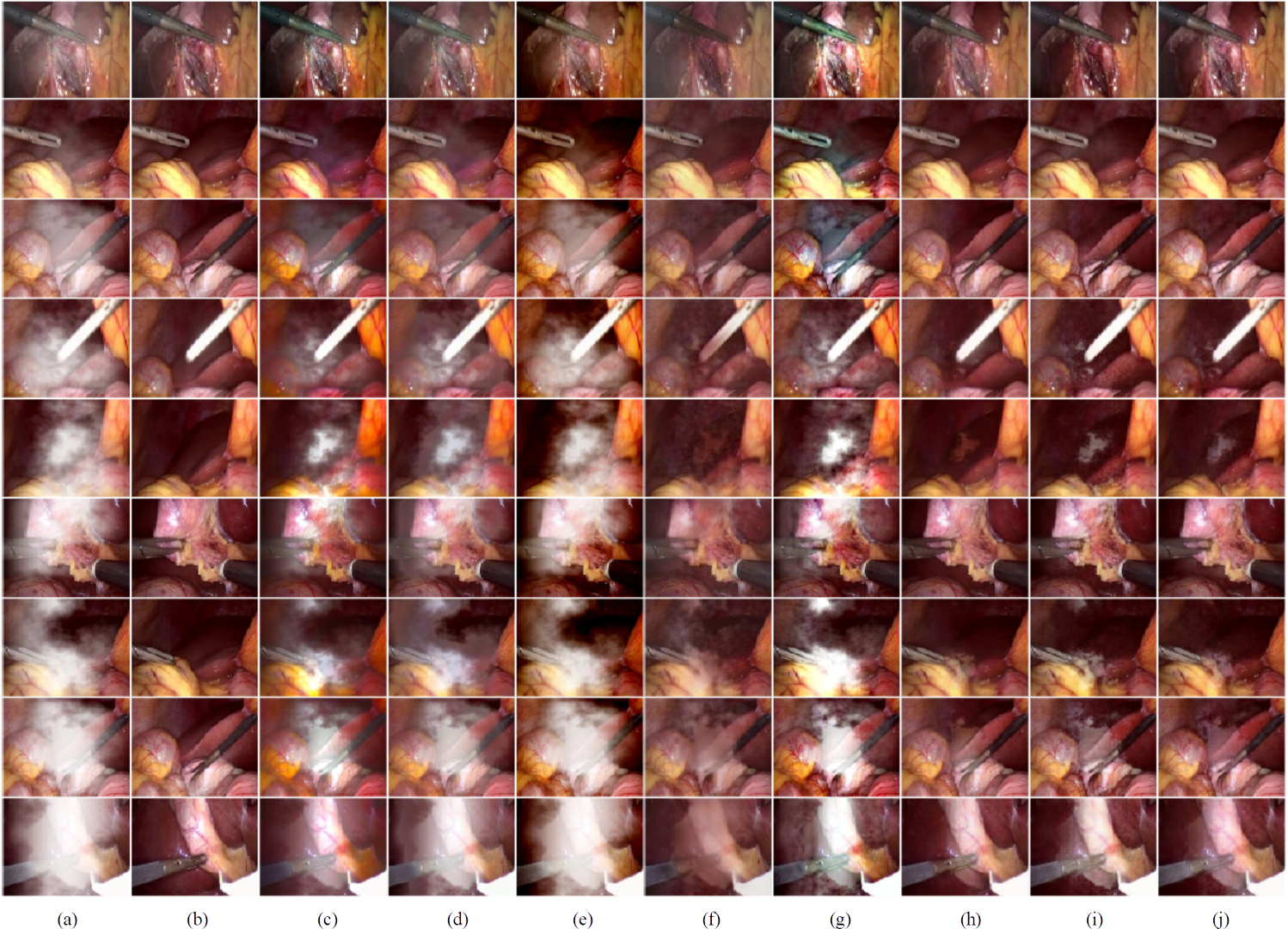}
    \caption{Comparison with the proposed methods in synthetic laparoscopic images. (a) Input images with synthetic smoke,  (b) ground-truth, (c) Pang et al. \cite{pang2011improved}, (d) Zhu et al. \cite{zhu2015fast}, (e) Bolkar et al. \cite{bolkar2018deep}, (f) Chen et al. \cite{chen2018unsupervised}, (g) Shin et al. \cite{shin2019radiance}, (h) Wang et al. \cite{wang2019multiscale}, (i) Isola et al. \cite{isola2017image}, and (j) the proposed method.}
    \label{synthetic}
\end{figure*}

It has been conducted several experiments comparing our results with the works cited in the previous Subsection. The results are summarized in Fig. \ref{synthetic}. From left to right it is shown in (a) nine input images with synthetic $CO_2$; in (b) their respective ground-truth, while (c) corresponds to the results of Pang et al. \cite{pang2011improved}; (d) to the results of Zhu et al. \cite{zhu2015fast}; (e) to the results of Bolkar et al. \cite{bolkar2018deep}; (f) to the results of Chen et al. \cite{chen2018unsupervised}; (g) to the results of Shin et al. \cite{shin2019radiance}; (h) to the results of Wang et al. \cite{wang2019multiscale}; (i) represents the output of the Pix2pix \cite{isola2017image}; and finally, (j) is the output of our proposed method. 

By making a visual inspection, it can be seen easily that Bolkar et al. \cite{bolkar2018deep} (Fig. \ref{synthetic}e), Shin et al. \cite{shin2019radiance} (Fig. \ref{synthetic}g), and Zhu et al. \cite{zhu2015fast} (Fig. \ref{synthetic}c) present the lowest performance in comparison with the rest of the works. On the other hand, Chen et al. (Fig. \ref{synthetic}h), and Isola et al. \cite{isola2017image} (Fig. \ref{synthetic}i), achieved a good reduction of synthetic gas, however, the color is affected in the output images. In the results of our method, shown in Fig. \ref{synthetic}j), the images are nearly close to the ground-truth since images preserve the original color and hold more details in the recovered zones. These aforementioned results are consistent with the quantitative analysis presented in Section \ref{QC} and depicted in Figs. \ref{boxplot1} and \ref{boxplot2}. 

\begin{figure*}
    \centering
    \includegraphics[width=0.98\textwidth]{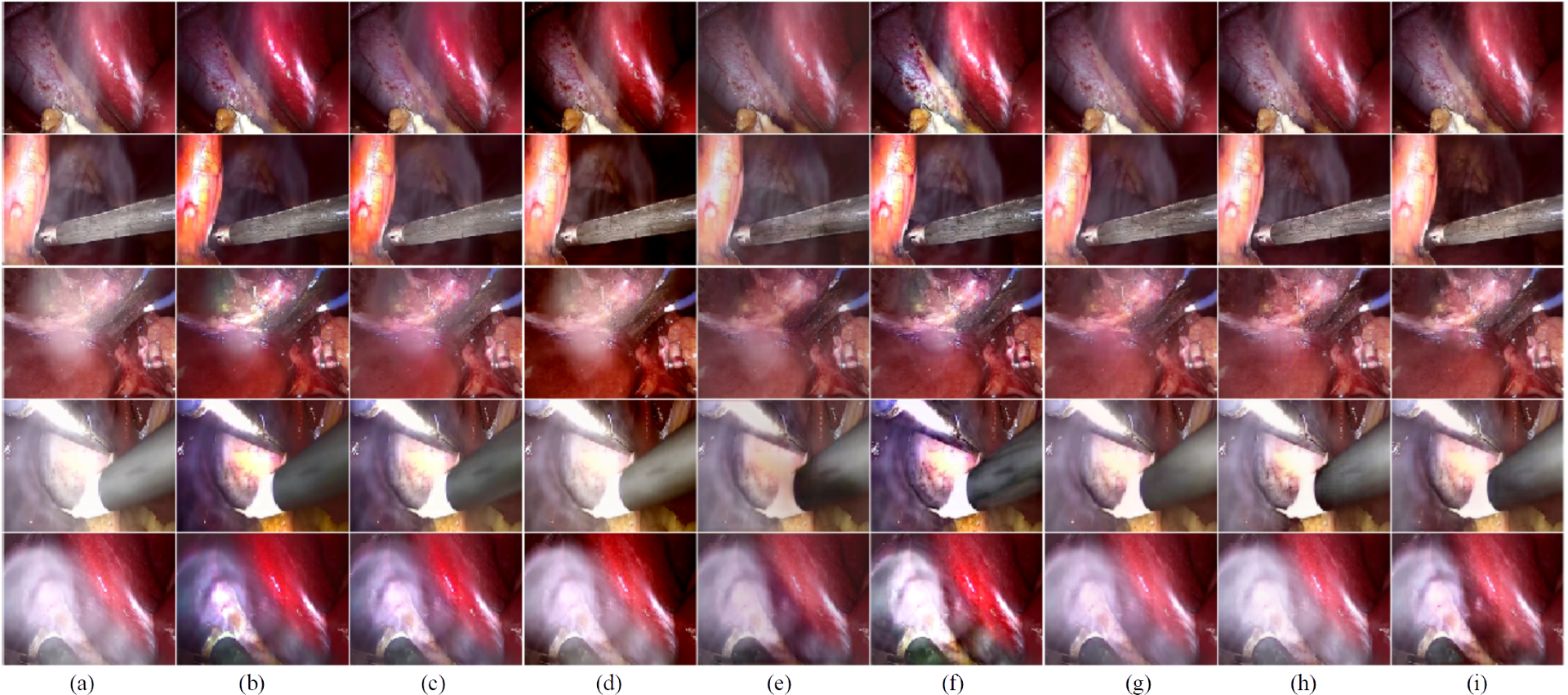}
    \caption{Comparison with the proposed methods in real laparoscopic images.  (a) Input images with smoke, (b) Pang et al. \cite{pang2011improved}, (c) Zhu et al. \cite{zhu2015fast}, (d) Bolkar et al. \cite{bolkar2018deep}, (e) Chen et al. \cite{chen2018unsupervised}, (f) Shin et al. \cite{shin2019radiance}, (g) Wang et al. \cite{wang2019multiscale}, (h) Isola et al. \cite{isola2017image}, and (i) the proposed method.}
    \label{real}
\end{figure*}

Real laparoscopic smoked images are shown in Fig. \ref{real}; from left to right: (a) five input images with smoke; from (b) to (i) the outputs of the current state-of-the-art methods are depicted. For the first group, the methods of Pang et al. \cite{pang2011improved} (Fig. \ref{synthetic}b) and Zhu et al. \cite{zhu2015fast} (Fig. \ref{synthetic}c) seem to be underperforming, since much of the gas can be seen in the output images. The method of Bolkar et al. \cite{bolkar2018deep} (Fig. \ref{synthetic}d) seems to reduce the effect of the gas more than the preceding two, nevertheless, it presents a major color saturation where darker regions look totally black.  For the second group, Chen et al. \cite{chen2018unsupervised} (Fig. \ref{synthetic}e) show a slight improvement over the methods of the first group, however, it presents opaque colors and poorly defined textures. In Shin et al. \cite{shin2019radiance} (Fig. \ref{synthetic}f), details and shapes are best appreciated, but colors present some saturation.
Wang et al. \cite{wang2019multiscale} (Fig. \ref{synthetic}g) and Pix2pix \cite{isola2017image}, (Fig. \ref{synthetic}h) show satisfactory results in terms of gas reduction, color and object detail. Finally, our method (Fig. \ref{synthetic}i) presents a minor gas influence of all the images, preserving fine details as well as the original colors.

\subsection{Quantitative comparison}\label{QC}
The results are evaluated using two metrics for image quality analysis commonly used in literature: the peak signal-to-noise ratio (PSNR), and the structural similarity index (SSIM). 

PSNR is a quantitative measure of the restoration quality and is defined as
\begin{equation}\label{eq7}
\textrm{PSNR}=10 \log_{10}\Bigl(\frac{\textrm{MAX}^2_I}{\textrm{MSE}}\Bigr)=20 \log_{10}\Bigl(\frac{\textrm{MAX}_I}{\sqrt{\textrm{MSE}}}\Bigr),
\end{equation}
where $\textrm{MAX}=2^B-1$; $B$ is the number of bits used in the image; and MSE is the Mean Square Error being its metric, which for two monochrome images $I$ and $J$ of size $m\times n$, is given by
\begin{equation}\label{eq6}
\textrm{MSE}=\frac{1}{mn}\sum_{x=0}^{m-1}\sum_{y=0}^{n-1}||I(x,y)-J(x,y)||^2.
\end{equation}
High values of PSNR indicate better restorations. The MPEG committee establishes an informal threshold of $\textrm{PSNR} = 0.5$ dB to decide whether a PSNR value is significant, and hence, differences in quality are visible \cite{salomon2004data}.

\begin{figure}[h]
    \centering
    \includegraphics[width=0.49\textwidth]{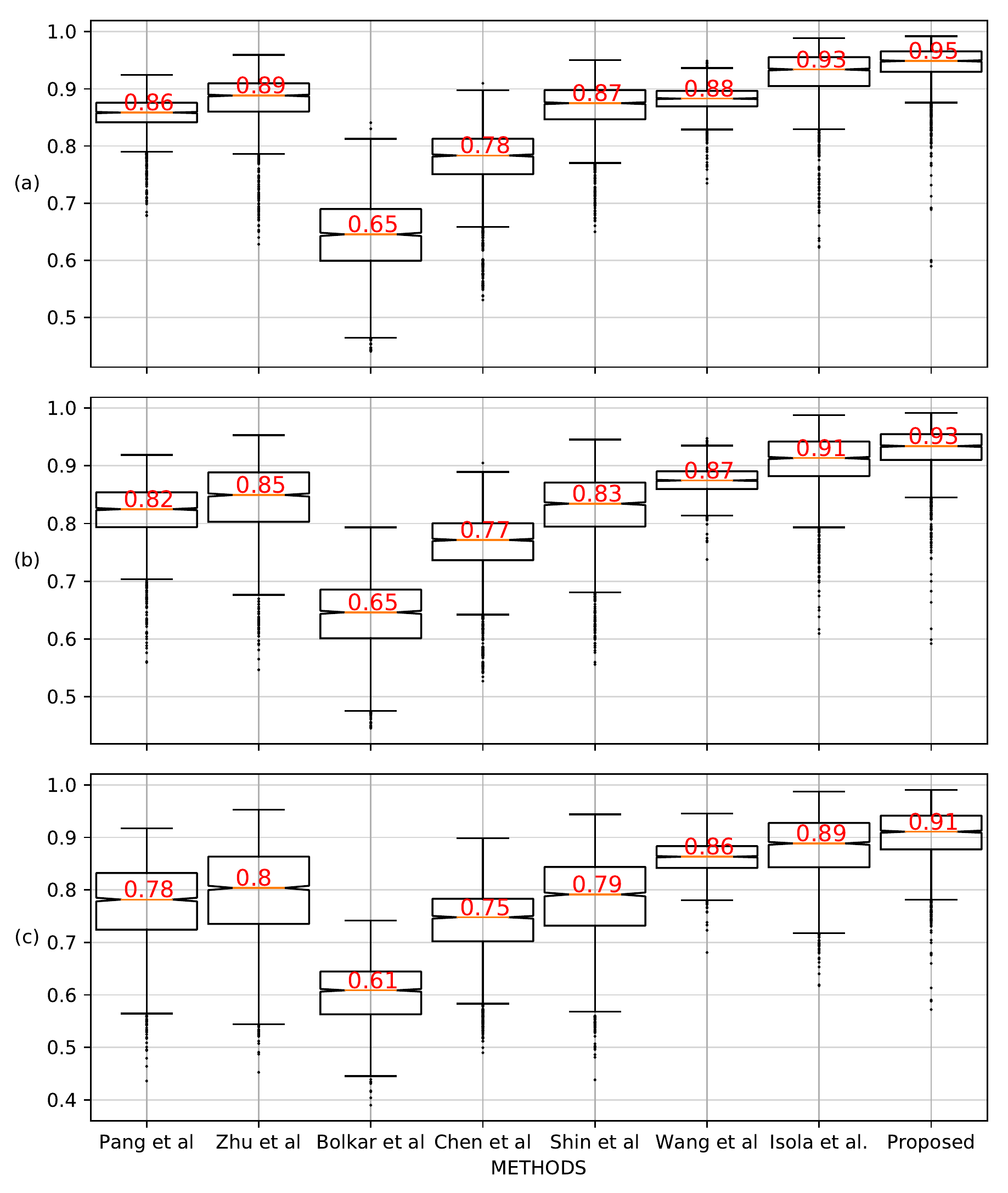}
    \caption{Performance comparison of the state-of-the-art methods according to the SSIM index on different smoke intensities: (a) low, (b) medium, and (c) heavy.}
    \label{boxplot1}
\end{figure}

\begin{figure}[h]
    \centering
    \includegraphics[width=0.49\textwidth]{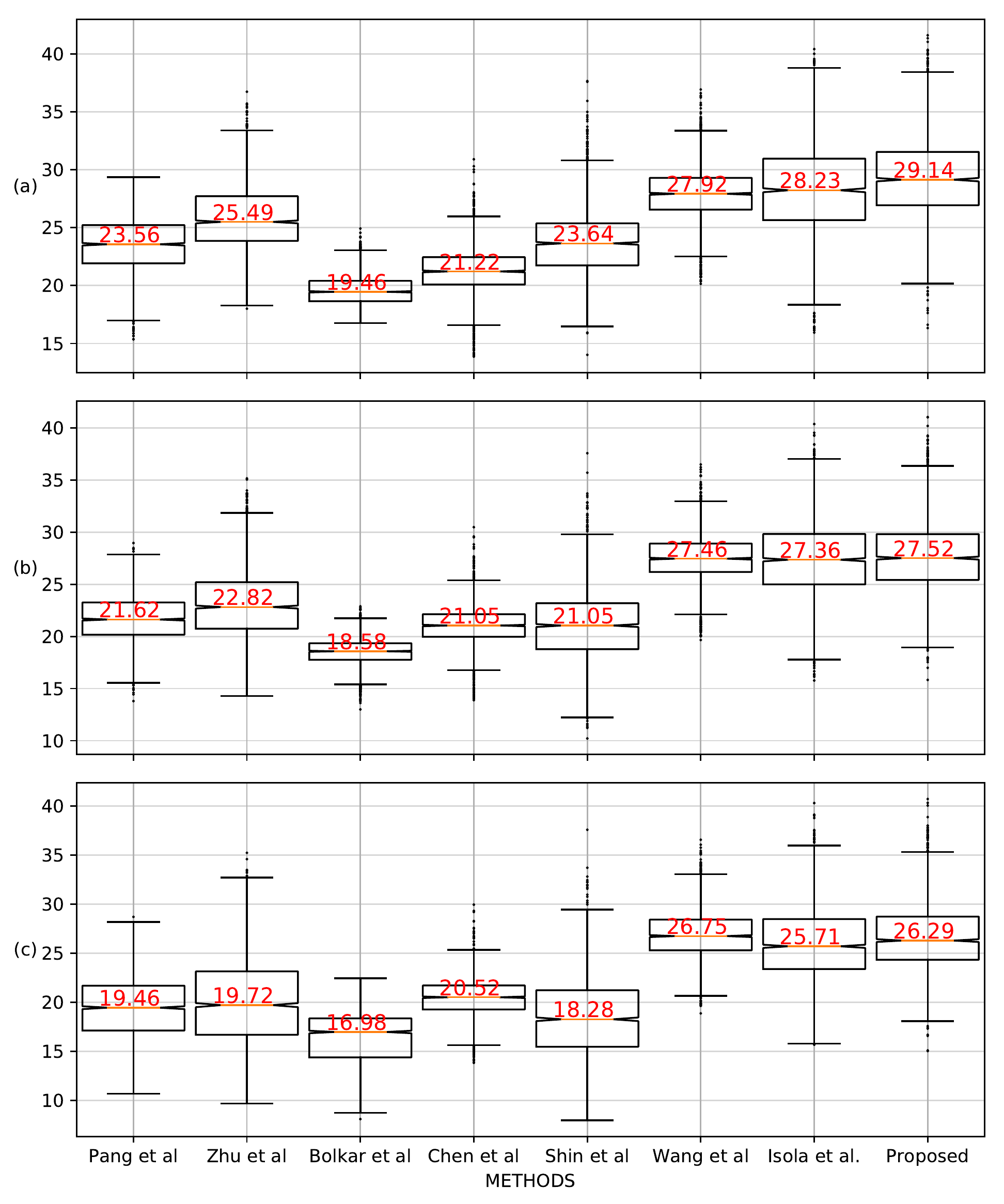}
    \caption{Performance comparison of the state-of-the-art methods according to the PSNR index on different smoke intensities: (a) low, (b) medium, and (c) heavy.}
    \label{boxplot2}
\end{figure}

SSIM index is a perceptual image similarity metric proposed as an alternative to MSE and PSNR indexes to increase the correlation with the subjective assessment. It is defined between $-1$ and $1$, where $-1$ represents a total anti-correlation, $0$ no correlation, and $1$ a total correlation between to images.
For the original and reconstructed images $I$ and $J$, SSIM index is defined as follows
\begin{equation}\label{eq8}
\textrm{SSIM}(I,J)= \frac{(2\mu_I\mu_J+C_1 )(2\sigma_{IJ}+C_2 )}{(\mu_I^2+\mu_J^2+C_1 )(\sigma_I^2+\sigma_J^2+C_2 ) },
\end{equation}
where $\mu$ is the mean; $\sigma$ is the variance; $\sigma_{IJ}$ is the covariance of the images; and $C_1, C_2$ are two variables that avoid the possibility of the denominator or numerator becoming zero. In Figs. \ref{boxplot1} and \ref{boxplot2} there are shown the SSIM and PSNR metrics for three different densities of synthetic $CO_2$: (a) is for low density; (b) is for medium density; and (c) is for high density. It can be observed that our proposed method presents the highest values for both metrics, except in case (c). In such a case the algorithm developed by Wang et al. \cite{wang2019multiscale} presents a PSNR value of 26.75, while our method has a value of 26.29, i.e., a difference of 0.46 below the 0.5 dB mentioned in \cite{salomon2004data}, whereby this difference is considered unimportant. In other cases, the difference in the PSNR metric in comparison with the rest of the methods is greater than 0.5. It is highlighted the dispersion reduction for the three smoke intensities presented in our method w.r.t. the pix2pix results. This can be seen in the metrics of SSIM and PSNR shown in Figs. \ref{boxplot1} and \ref{boxplot2}, respectively.
%

\subsection{Processing time comparison}
Since the main goal is to support surgeons during laparoscopic surgery, the processing time is crucial. Table \ref{tabla3} shows the compared methods and their respective average processing time in frames per second (fps). It is highlighted that our method has not been optimized. Since our method computes the dark channel, the processing time is a little higher than the pix2pix, in which evidently the dark channel is not computed. This is shown in Table \ref{tabla3}. Also, it is noted in Table \ref{tabla3} that our method has a better performance than the state-of-art methods. Since in the experimental platform our method outperforms by three times the common video frame rate of 30 fps, it is possible to implement our algorithm in a real-time system, and even in a low-cost device. In \footnote{\href{https://youtu.be/Gw8OZNDdicE}{Link to a demonstrative video of our method.}} it is shown a demonstrative video of the proposed method in a real-world laparoscopic images.

\begin{center}
\begin{table}[h]
\renewcommand{\arraystretch}{1.6}
\caption{Processing time comparison in frames per second (fps).}
\label{tab:time}
\centering
\tiny	
\resizebox{0.49\textwidth}{!}{
\begin{tabular}{lcl}
\hline
\textbf{Methods} & \textbf{Processing time (Frames/Sec)} & \textbf{Platform}  \\ \hline\hline
Pang et al.   \cite{pang2011improved}                   & 2.98          & Matlab     \\ \hline
Zhu et al.    \cite{zhu2015fast}                        & 16.60         & Matlab     \\ \hline
Bolkar et al. \cite{bolkar2018deep}                     & 32.40         & Python (Caffe)     \\ \hline
Chen et al.   \cite{chen2018unsupervised}               & 89.14         & Python (Tensor Flow)    \\ \hline
Shin et al.   \cite{shin2019radiance}                   & 1.28          & Matlab     \\ \hline
Wang et al.   \cite{wang2019multiscale}                 & 24.00         & Python (Keras)     \\ \hline
Isola et al.  \cite{isola2017image}        & 120.0         & Python (Pytorch)    \\ \hline
Our proposed method & 92.19         & Python (Pytorch)    \\ \hline
\end{tabular}}
\label{tabla3}
\end{table}
\end{center}
\newpage
\section{Conclusions}\label{sec:conc}
The lack of visibility on surgical procedures, caused by the smoke of $CO2$ and laparoscopic cautery, increases the possibility of errors and time surgery. Therefore, a method to remove the smoke effects is necessary to support the surgeons, thus reducing risks for the patient and increasing the efficiency of medical specialists. In this work, it is proposed the use of the dark channel embedded in the input of a generator network. 

The addition of dark-channel-guide contributes to the identification of the regions with the presence of smoke, focusing on the restoration of those regions. Qualitative evaluations in synthetic and real images showed that this new approach generates output images with better contrast and color preservation than the other seven state-of-the-art methods. On the other hand, quantitative evaluations based on PSNR and SSIM indexes for three different densities of synthetic smoke, showed that the proposed approach outperformed all the seven aforementioned algorithms. Moreover, the processing time of 92.19 frames per second shows that the proposed architecture can be easily implemented into a system for real-time applications used in medical devices. 

\section*{ACKNOWLEDGMENT}
Hugo Alberto Moreno especially thanks to CONACYT (National Council for Science and Technology) for the financial support provided for his master's studies.
\ifCLASSOPTIONcaptionsoff
  \newpage
\fi


\bibliography{references}
\bibliographystyle{ieeetr} 

\end{document}